# Néel-type skyrmion lattice in tetragonal polar magnet VOSe$_2$O$_5$

Takashi Kurumaji[1]*, Taro Nakajima[1], Victor Ukleev[1], Artem Feoktystov[2],

Taka-hisa Arima[1,3], Kazuhisa Kakurai[1,4], and Yoshinori Tokura[1,5]

[1] *RIKEN Center for Emergent Matter Science (CEMS), Wako 351-0198, Japan*

[2] *Jülich Centre for Neutron Science (JCNS) at Heinz Maier-Leibnitz Zentrum (MLZ), Forschungszentrum Jülich GmbH, Garching 85748, Germany*

[3] *Department of Advanced Materials Science, The University of Tokyo, Kashiwa 277-8561, Japan*

[4] *CROSS-Tokai, Research Center for Neutron Science and Technology, Tokai, Ibaraki 319-1106, Japan*

[5] *Department of Applied Physics, The University of Tokyo, Tokyo 113-8656, Japan*

*Corresponding author: takashi.kurumaji@riken.jp

**Abstract**

Formation of the triangular skyrmion-lattice is found in a tetragonal polar magnet $VOSe_2O_5$. By magnetization and small-angle neutron scattering measurements on the single crystals, we identify a cycloidal spin state at zero field and a Néel-type skyrmion-lattice phase under a magnetic field along the polar axis.  Adjacent to this phase, another magnetic phase of an incommensurate spin texture is identified at lower temperatures, tentatively assigned to a square skyrmion-lattice phase. These findings exemplify the versatile features of Néel-type skyrmions in bulk materials, and provide a unique occasion to explore the physics of topological spin textures in polar magnets.

Skyrmions have been investigated in various magnetic systems [1-5], especially noncentrosymmetric magnets, where the topological spin textures are stabilized by Dzyaloshinskii-Moriya (DM) interaction due to the relativistic spin-orbit coupling [6-9] under an applied magnetic field ($H$). A periodic lattice of skyrmions is typically found in chiral magnets such as B20 alloys [1,2,10], multiferroic $Cu_2OSeO_3$ [11], and *β*-Mn-type CoZnMn alloys [12], where the skyrmion form Bloch-type, whirl-like spin vortex structure. Early theoretical predictions [8,9] and subsequent experiment [13], however, revealed that a different type of skyrmion emerges due to another kind of asymmetry of the underlying lattice. Polar systems exemplify one such noncentrosymmetry, in which DM interaction confines the magnetic modulation direction vector ($\vec{q}$) perpendicular to the polar axis to stabilize the cycloidal spin order, as shown in Fig. 1(a). In this case, $H$ applied parallel to the polar (*c*) axis induces the Néel-type skyrmion as shown in Fig. 1(b): the spin rotates outwards from the core of the vortex.

The Néel-type skyrmion has been frequently observed in magnetic ultra-thin films and multilayers affected by DM interaction via broken inversion symmetry at the interface [3,4,14-16], which provide an important arena for practical applications in spintronics devices. In this context, a precise understanding of Néel-type skyrmions and related spin textures by targeting a bulk polar magnet is of potential importance as the foundation for skyrmion device research. However, the skyrmion lattice (SkL) phase in polar bulk magnets has not been fully investigated, except for a lacunar spinel compound [13] with trigonal crystal structure. In order to gain insights on the physical origin and the stability of skyrmion and related spin textures in a noncentrosymmetric magnet, expansion of material classes of bulk polar magnets hosting Néel-type SkL is desired.

We have investigated the tetragonal polar magnet, $VOSe_2O_5$, which belongs to the $C_{4v}$ point group satisfying the prerequisite for hosting the Néel skyrmion [8]. Its crystal structure (Fig. 1(c)) consists of stacked square lattices of $VO_5$ tetragonal pyramids [17], each of which carries a magnetic $V^{4+}$ ion with spin-1/2 moment. Ferromagnetic like ordering at Curie temperature ($T_C$) ~8 K was reported [18], and the subsequent powder neutron diffraction investigation with density functional calculation [19] suggested a 3-up-1-down ferrimagnetic spin order in the unit cell, as shown in Fig. 1(d); this model predicts spontaneous magnetization 0.5 $\mu_B$/f.u., compared with the observed field-induced saturation, ~0.47 $\mu_B$/f.u., at the lowest temperature [19]. Some anomalies in magnetic susceptibility have remained unidentified, implying nontrivial magnetism in this compound [18,19].

In this single-crystal study, we performed measurements of magnetization and small-angle neutron scattering (SANS) to unveil the magnetic phases, including a triangular SkL phase under $H\|c$, as well as the cycloidal spin order propagating perpendicular to the polar axis in zero field. We found that the triangular SkL competes with versatile magnetic phases, one of which is distinct from the cycloidal or collinear spin state but shows magnetic modulation with the four-fold symmetric SANS pattern under $H\|c$. We compared our results with the recently discovered Néel-SkL in trigonal $GaV_4S_8$ [13] to gain insights on the stable topological spin textures possible in a polar magnet.

Single crystals of VOSe$_2$O$_5$ were grown by chemical vapor transport reaction with NH$_4$Cl [20]. Sizes of the obtained single crystals were typically 0.5 × 0.5 × 0.7 mm$^3$. Magnetization and AC magnetic susceptibility were measured by a superconducting quantum interference device magnetometer (MPMS3, Quantum Design). To measure AC susceptibility, typically a 1 Oe AC field at 100 Hz was applied along the DC $H$ direction. During the $H$ scan measurement, such as in Fig. 2, the temperature precision is better than ±0.01 K. SANS measurements were performed at the KWS-1 beamline at Heinz Maier-Leibnitz Zentrum (MLZ), Garching, Germany [20-22]. We employed twenty-seven pieces of crystals (total volume of 4.6 mm$^3$), which were carefully co-aligned on an aluminum plate with the same crystallographic orientation [20]. The sample on the Al-plate was mounted into a $^3$He-circulation refrigerator with its [001] direction parallel to the incident beam. A magnetic field parallel or perpendicular to the incident beam was generated by an electro-magnet. To avoid the effect of residual fields, the electromagnet was demagnetized before performing subsequent zero/applied field experiments.

In the SANS pattern (Fig. 1(e)) for the co-aligned single crystals at 6.2 K in zero field with the incident neutron beam parallel to the $c$ axis, we observed in-plane magnetic modulation. Four clear magnetic Bragg reflections are observed at $q = 0.046$ nm$^{-1}$ along the crystallographically equivalent $a$ and $b$ axes, suggesting the multidomain nature of the single-$q$ spin state for $\vec{q}||\vec{a}$ and $\vec{q}||\vec{b}$. We further find that a single-$q$ domain can be selected with in-plane $H$. Figure 1(f) shows the zero-field SANS pattern at 6.3 K after a field-trained procedure with $H = 120$ Oe along the $a$ axis [20]. Two out of four Bragg reflections satisfying the $\vec{q} \perp \vec{H}$ configuration are observed, which is consistent with the anticipated cycloidal spin structure [23]. This result indicates that the spin modulation in this system obeys the DM interaction for the polar symmetry, which determines the Néel-type spin configuration as the stable skyrmion form under $H//c$.

The magnetic transition from the cycloidal spin phase to the SkL phase, or the so-called A-phase, under $H \| c$ is exemplified by the two-step metamagnetic transition in magnetization ($M$) at 7.45 K (Fig. 2(a)), near $T_C$ = 7.50 K. This is a common feature of SkL formation, as observed in various skyrmion-hosting compounds [11,13,24,25]. We identify the A-phase boundary by the double-peak structure in both the real ($\chi'$) and imaginary ($\chi''$) parts of the AC magnetic susceptibility (Fig. 2(b)). This $\chi'$-valley region disappears at slightly lower temperature ($T < 7.20$ K). However, at 6.00 K (Figs. 2(c)-2(d)), $H$-dependence of $M$ and $\chi'$ still show nonmonotonous behavior around 15 Oe. This phase transition is of first-order nature, as the peak for $\chi''$ suggests dissipation due to the motion of a domain wall separating the cycloidal state and distinct magnetic states, IC-1 and IC-2, respectively (top abscissa in Fig. 2(c)). Here, the notation, IC, stands for an incommensurate magnetic order, which is confirmed by the SANS investigation as mentioned earlier and below. Figure 2(e) is the magnetic phase diagram for $H \| c$, determined from the magnetization measurement, with the color plot of $\chi'$. A pocket-like magnetic phase around $T_C$ with a finite field (as indicated by A) is clearly identified, as well as the IC-2 phase in a lower temperature region.

The triangular SkL formation in the A-phase is identified by a twelve-fold SANS pattern in the configuration with $H$ parallel to both the $c$ axis and the neutron beam.

Figure 3(a) shows the SANS pattern at 7.45 K with $H$ = 25 Oe along the $c$ axis. Note here that the A-phase region for the SANS sample exists in a higher $H$ region than that for Fig. 2(a) due to the sample-shape-dependent difference of demagnetization fields; see rescaled $H$ at the right ordinate in Fig. 2(e) [20]. We took a constant-field path with decreasing temperature from the paramagnetic state to the A-phase. Besides four Bragg reflections along the $a$ and $b$ axes, additional spots show up (Fig. 3(a)) at the positions irrelevant to the tetragonal symmetry. Schematic SANS pattern is shown in Fig. 3(b), which can be addressed to the superposition of three SANS patterns as Figs. 3(c)-3(e). Among them, Figures 3(c) and 3(d) show two types of the triangular SkL state with 90° rotation from each other, in which one $q$ of the triple-$q$ structure is fixed along the $a$ (Fig. 3(c)) or $b$ (Fig. 3(d)) axis, respectively. As shown by the azimuthal angle ($\phi$) dependence of the integrated intensity in Fig. 3(f), the separated spots with 30° period clearly reflect the hexagonal symmetry of the superposing spin textures. Similar twelve-fold SANS patterns due to a multidomain state of the triangular SkL have been reported in other skyrmion-hosting materials such as CoZnMn alloy [26] and $Cu_2OSeO_3$ [27] under $H$ along [001] axis, similarly to the present case. Superposition with a four-fold pattern (Fig. 3(e)) appears to be due to the coexisting IC-1 and/or IC-2 states because the assembled single crystals for the SANS investigation effectively experience an inhomogeneous magnetic field due to the different demagnetization effect for individual pieces.

It is confirmed that further cooling destructs this twelve-fold SANS pattern for the triangular SkL state. As shown by the red circle in Fig. 3(g), when cooled to 6.00 K with $H$ = 25 Oe (Fig. 3(g)), the twelve-fold intensity profile evolves into four peaks with 90° period, in accord with the formation of the IC-1 and IC-2 phase domains. Note that weak peaks, indicated by asterisks in Fig. 3(g), involving the triangular SkL are observed at 6.00 K. This suggests that a small portion of the triangular SkL state for Fig. 3(d) remains in a super cooled metastable state, as has been identified in other SkL-materials [26,28,29]. The remnant SkL peaks were not observed in the SANS pattern for $H$ = 20 Oe (or 30 Oe) at 6.00 K in the $H$-increasing process after the zero-field cooling (ZFC) procedure (Fig. 3(g)), which suggests the metastability of these remnant SkL structures. Note that the SANS measurement offers the reciprocal-space image of the spin texture. Further investigation by the real space observation technique is desired for more concrete proofs of the spin texture such as the two-domain state of the SkLs and the quenched SkL state at lower temperatures coexisting with IC-1/IC-2 states.

To connect the A-phase region identified by the magnetic susceptibility measurement and the twelve-fold SANS pattern, we performed further analysis on the SANS intensity for each spin state. We partitioned the reciprocal plane as shown in Fig. 3(b) to plot the temperature dependence of each integrated intensity in Fig. 3(i): the intensity for the four-fold pattern ($I_4$) and that of the diagonal region ($I_6$) representing contribution originating from the triangular SkL. Here, $I_6$ is scaled by 3/2 to compare the intensity of individual peaks. Both $I_4$ and $I_6$ increase as the temperature decreases in accord with the thermal evolution of the spin moment. The intensity in the $I_6$ region originates from the following two components: (1) the broadened Bragg spots in the $I_4$ region due to flexibility of the $q$ vector for the in-plane direction; (2) the super cooled triangular SkL state. To renormalize the thermal increase of $I_6$, we plotted the relative intensity $I_6/I_4$ for the right ordinate in Fig. 3(i). $I_6/I_4$ peaks in the triangular SkL region between χ' peaks (Fig. 3(h)), pink hatching, then decays with decreasing temperature to a finite value. These results support that the A-phase region of the phase diagram is occupied by the thermally equilibrium triangular SkL state.

Note that the presence of the IC-2 phase is unique in the present tetragonal system; in the trigonal Néel-skyrmion material $GaV_4S_8$, the triangular SkL phase is in proximity with the cycloidal, ferromagnetic, and paramagnetic phases [13]. A typical SANS pattern for the IC-2 phase taken at 6.00 K with $H$ = 50 Oe along the $c$ axis after the ZFC procedure is shown in Fig. 4(a) (see the rescaled $H$ in Fig. 2(e)). Clear four-fold Bragg reflections along the $a$ and $b$ axes can be observed. To identify the magnetic transition between IC-1 and IC-2 states, we show the $H$-dependence of the integrated intensity and wavelength ($\lambda$) of the magnetic modulation together with χ' and χ'' measured for the assembled-crystals sample (SANS sample) (Figs. 4(b)-4(d)). Although the integrated intensity monotonously wanes towards the ferrimagnetic (FM) phase, $\lambda$ shows a discernible kink at ~40 Oe as indicated by a triangle (Fig. 4(d)), which correlates with features in χ' and χ'' (Fig. 4(b)). This suggests that the IC-2 phase has in-plane magnetic modulation while being distinct from the cycloidal spin order (IC-1), or any other spin order such as a chiral soliton lattice state [30] continuously evolved from IC-1 under $H||c$.

Enriched magnetic phase diagram for this system is also exemplified by the thermal evolution of the SANS integrated intensity for the IC-1 state in zero field as shown in Fig. 4(e). The intensity increases as the temperature decreases from $T_C$ until a collapse occurring at ~4.5 K. This successive transition was previously suggested by the AC susceptibility measurement using powder sample [19]. We measured the temperature dependence of $\chi'$ in zero field using the single crystal for the AC magnetic field ($H_{ac}$) parallel to the $c$ axis (Fig. 4(e)). Note that the disappearance of SANS intensity does not correlate with the peak for $\chi'$ at 4.0 K but with the anomaly at 4.5 K as indicated by a black triangle in Fig. 4(e). The hysteresis in the temperature dependence of the SANS intensity and $\chi'$ around 5 K indicates the presence of the other magnetic phase (B-phase) between the IC-1 and the easy-plane FM ground state. The nature of the magnetic structure of B-phase is not fully clear, but a plausible candidate is a commensurate canted ferrimagnetic order ($q = 0$), which may be derived by the diverging tendency of $\lambda$ as observed in Fig. 4(f) prior to entering the ground state. Figure 4(g) summarizes the $H$-$T$ phase diagram. IC-1 and IC-2 are restricted to a limited temperature range (4.0 K $< T <$ 7.5 K), due to the transition into the easy-plane FM state and B-phase at lower temperatures. The saturation magnetic field under $H\|c$ (Fig. 4(g), open squares) rapidly increases with lowering temperature, which further supports the growing easy-plane anisotropy.

The increase in $\lambda$ for IC-1 with decreasing temperature (Fig. 4(f)) suggests the thermal variation of the magnitude of symmetric and antisymmetric (DM) exchange interactions, and the magnetic anisotropy of the present multi-sublattice system. The role of such parameters for stable spin textures in a polar system under a magnetic field has been discussed with various computational approaches [31-34]. Several theoretical studies predicted that a polar magnet, in contrast to a chiral magnet, could have versatile magnetic phases, including the square SkL state under $H$, which is more stable than the triangular SkL state in the presence of easy plane anisotropy [31,33,34]. Note that the tetragonal anisotropy may also be relevant in the present system since the *q* vector for the cycloidal spin order is weakly locked along the *a* or *b* axis (Fig. 1(e)) in zero field, in contrast to the homogeneous ring observed at the SANS pattern in the case of the trigonal $GaV_4S_8$ [23]. Although the formation of a multiple-*q* state in the IC-2 phase remains elusive within the present experimental information, a square SkL state is a plausible candidate under the interplay between the thermally-changing uniaxial and tetragonal magnetic anisotropies originating from the underlying crystal structure. Further theoretical and experimental studies are needed to clarify the stable spin texture in this unique tetragonal polar magnet.

In conclusion, we experimentally identified the triangular skyrmion lattice (SkL) phase in a tetragonal polar system $VOSe_2O_5$. Cycloidal spin modulation in the zero field evidences the stability of the Néel-type skyrmion in the present system. We unraveled the relation between triangular SkL state and the other magnetic phase with in-plane modulation at low temperature under $H||c$. This phase is different from the cycloidal spin order in zero field, and possibly a square SkL state induced by the interplay of the tetragonal crystal anisotropy and the effect of thermally-varying uniaxial anisotropy. This system shows a distinct magnetic correlation with the recently discovered trigonal polar skyrmion material, providing novel insights into the stabilization of SkL states in polar magnets.

**Acknowledgements**

The authors thank D. Hashizume for support for administration and operation of a single-crystal X-ray diffractometer; Dr. T. Reimann, M. Seifert, and Dr. M. Schulz for providing us the magnet and helping with its operation.   We also thank X. Z. Yu, D. Morikawa, M. Kriener, T. Matsumoto, Y. Tokunaga, I. Kezsmarki, and L. Ye for enlightening discussions. This research was supported in part by JSPS Grant-In-Aid for Scientific Research(S) No. 24224009, and Grant-In-Aid for Young Scientists(B) No. 17K14351.

**Figure Captions**

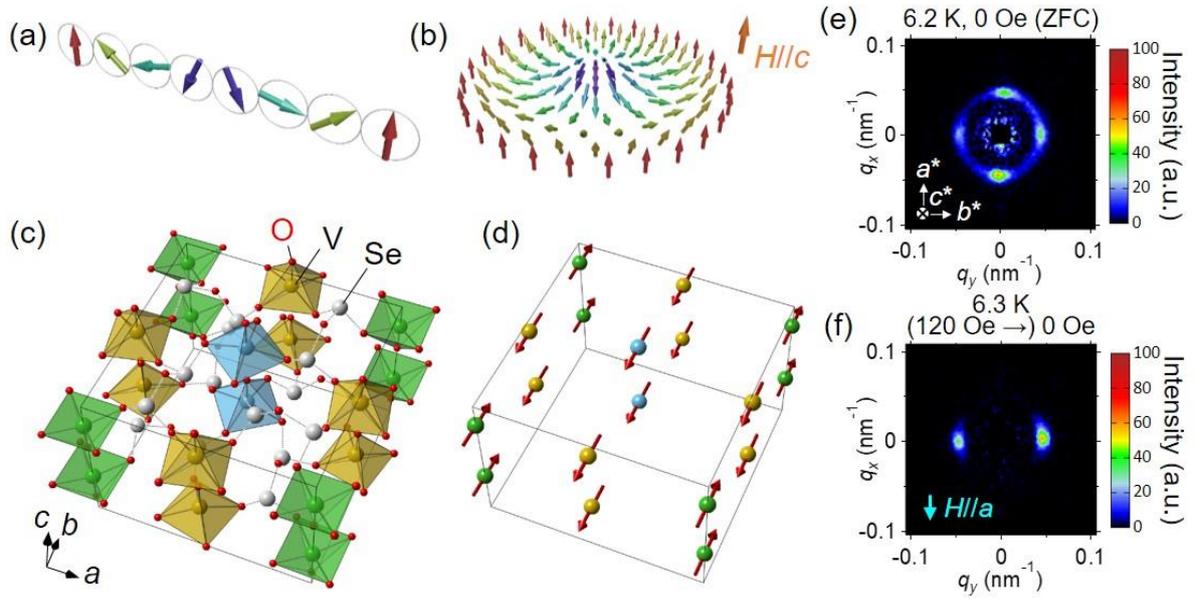

FIG. 1: Schematic spin configuration of (a) cycloidal spiral order and (b) a Néel-type skyrmion under magnetic field ($H$) along the $c$ axis. (c) Crystal structure of $VOSe_2O_5$. Blue, green, and yellow tetragonal pyramids are inequivalent $VO_5$ polyhedra in a unit cell. (d) Field-induced ferrimagnetic order in $VOSe_2O_5$. (e,f) Small-angle neutron scattering (SANS) pattern for the co-aligned single crystals in zero field at 6.2 K ($\pm 0.1$ K), (e) after the ZFC procedure, and (f) after the field-trained process with $H = 120$ Oe along the $a$ axis. $H$ is perpendicular to the incident neutron beam. The incident beam is parallel to the $c$ axis.

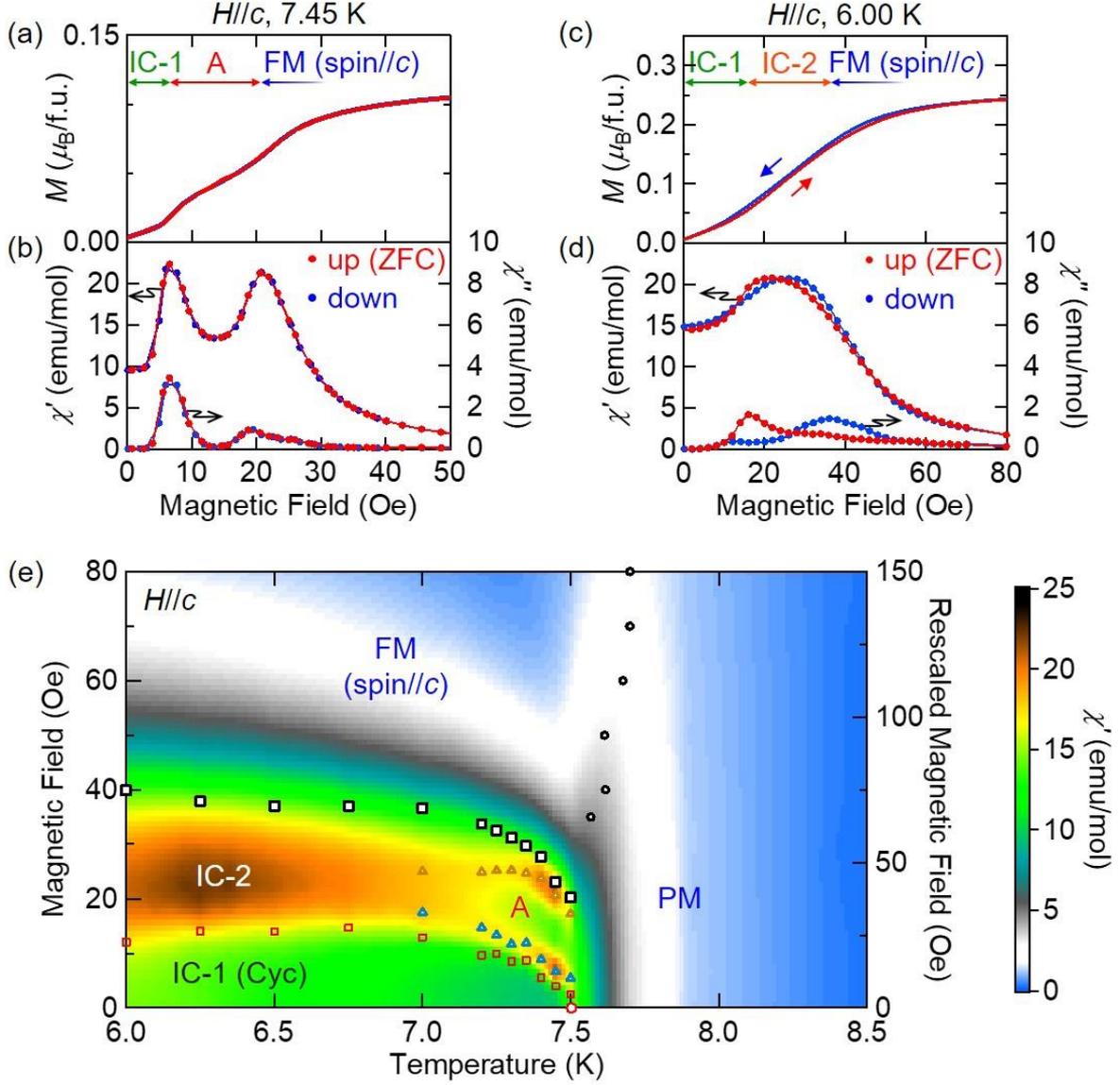

FIG. 2: (a-d) $H$ dependence of magnetization ($M$), and real ($\chi'$) and imaginary ($\chi''$) part of AC magnetic susceptibilities under $H//c$ at (a,b) 7.45 K, and (c,d) 6.00 K, respectively. Red (blue) circle is for $H$-increasing (-decreasing) scan. $H$-increasing scan was performed after the ZFC procedure. (e) Color plot of $\chi'$ for $H//c$, and the magnetic phase diagram determined by the measurements of $M$, $\chi'$, and $\chi''$, where each magnetic phase is indicated as PM: paramagnetic; FM: ferrimagnetic (3-up-1-down type state with spin//$c$); IC-1 (Cyc): cycloidal; IC-2: incommensurately modulated magnetic order (likely square skyrmion lattice state; see the text), and A: triangular skyrmion lattice state. Open circle: peak in $\chi'$-$T$ curve; open triangle: peak in $\chi'$-$H$ curve; and open square: peak in $\frac{d\chi'}{dH}$-*H* curve. Rescaled *H* for the phase diagram of the assembled-crystals sample for the SANS investigation is shown on the right ordinate.

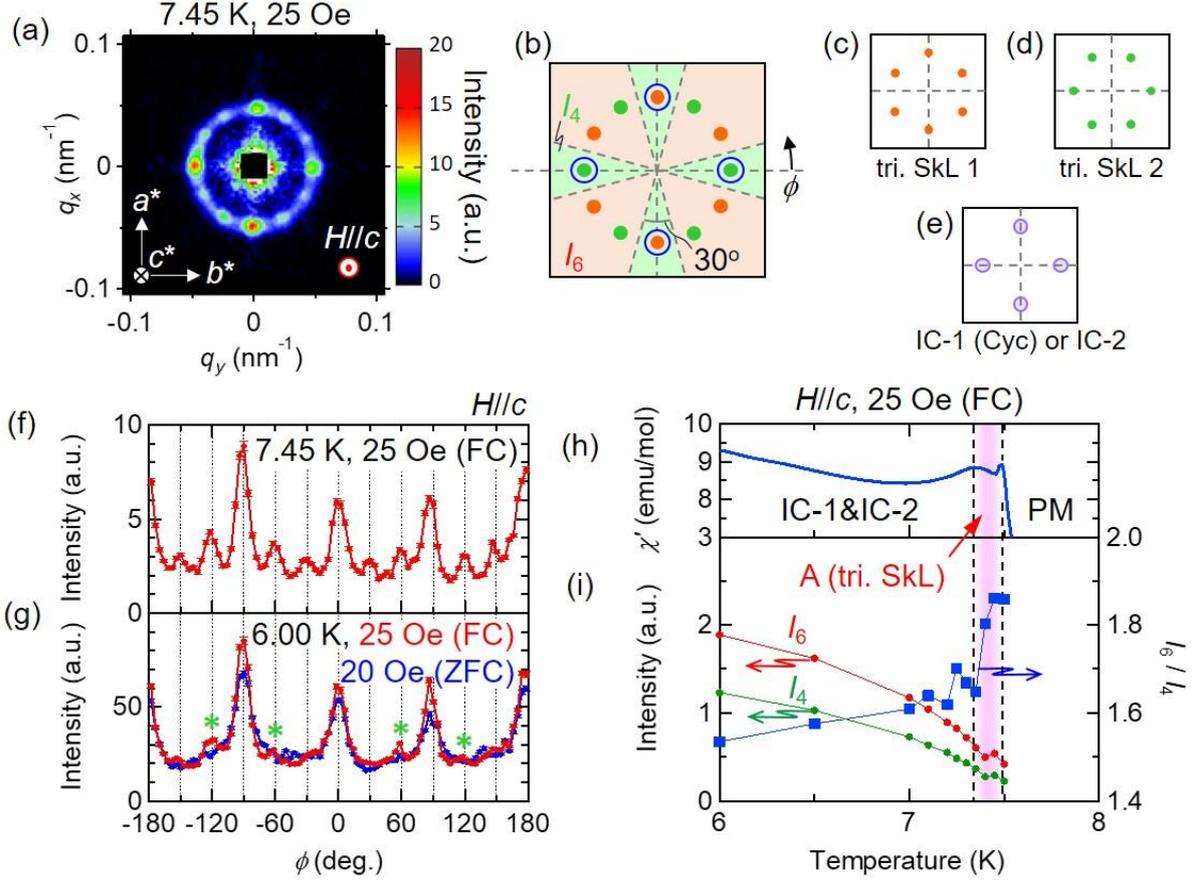

FIG. 3: (a) SANS pattern for the co-aligned single crystals at 7.45 K with $H =$ 25 Oe along the $c$ axis. $H$ and the incident neutron beam are parallel with each other and both along the $c$ axis. (b) Schematic illustration of the SANS pattern for (a). Blue, green and orange circles indicate the respective Bragg spots corresponding to the magnetic orders shown in (c), (d), and (e). Integration sectors for the intensity of magnetic four-fold spots (green region) $I_4$ and SkL (red region) $I_6$. Definition of the azimuth angle ($\phi$) is also indicated. (c-e) Schematics of SANS pattern for (c,d) two kinds of domain of triangular SkL and (e) four-fold pattern for multidomain of cycloidal order in IC-1 phase or for IC-2 phase. (f,g) $\phi$ dependence of integrated intensity for the field-cooling (FC) process with 25 Oe for $H//c$ at (f) 7.45 K, and (g) 6.00 K (red circle). In (g), the data with 20 Oe for $H//c$ after the ZFC procedure are also plotted by blue circles. The intensity integration is done over the radial $|\vec{q}|$ range $0.038 \leq |\vec{q}| \leq 0.056$ nm$^{-1}$ for 7.45 K, and $0.036 \leq |\vec{q}| \leq 0.053$ nm$^{-1}$ for 6.00 K. The remnant intensity peak for a super cooled component of the triangular SkL state is indicated by asterisk.   (h,i) Temperature dependence of (h) $\chi'$, (i) $I_4$, $I_6$, and their ratio $I_6/I_4$ for the field-cooling process with 25 Oe for $H//c$.

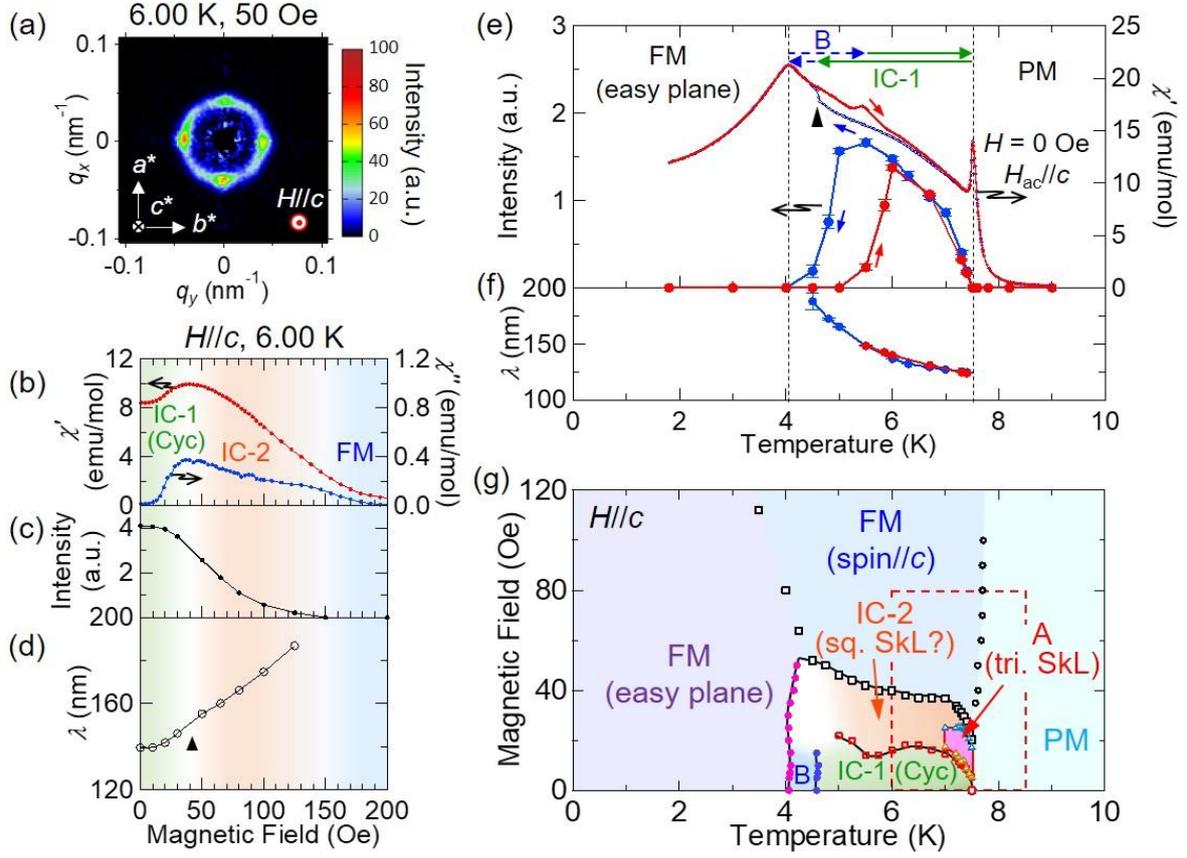

FIG. 4: (a) SANS pattern for the co-aligned single crystals at 6.00 K with 50 Oe for $H//c$. $H$ and the incident neutron beam are parallel with each other and both along the $c$ axis. (b-d) $H$ dependence of (b) $\chi'$ and $\chi''$, (c) integrated SANS intensity for overall detector surface, and (d) the wavelength $\lambda$ of the magnetic modulation. (e,f) Temperature dependence of the respective data measured in zero field for a cooling (blue circles) scan and a heating scan (red circles): (e) the SANS integrated intensity and $\chi'$ with the AC magnetic field $H_{ac}$ along the $c$ axis and; (f) $\lambda$ measured by the SANS pattern with the neutron beam parallel to the $c$ axis. (g) Magnetic phase diagram of $VOSe_2O_5$ determined by magnetization measurements with $H//c$. Phase boundaries for the white region (~4.5 K, ~30 Oe) could not be definitely determined by the magnetization measurements. For the definition of each symbol, see the figure caption for Fig. 2(e). Red dashed square indicates the $H$-$T$ region shown in Fig. 2(e).

Supplemental material for

**Néel-type skyrmion lattice in tetragonal polar magnet VOSe$_2$O$_5$**

Takashi Kurumaji[1]*, Taro Nakajima[1], Victor Ukleev[1], Artem Feoktystov[2],

Taka-hisa Arima[1,3], Kazuhisa Kakurai[1,4], and Yoshinori. Tokura[1,5]

1. **Single crystal growth and magnetization measurement with assembled single crystals.**

2. **SANS investigation with a magnetic field along the *a* axis.**

1. **Single crystal growth and magnetization measurement with assembled single crystals.**

Single crystals of $VOSe_2O_5$ were grown by the chemical vapor transport reaction. Sizes of the obtained single crystals were typically $0.5 \times 0.5 \times 0.7$ mm$^3$. First, powder of $VOSe_2O_5$ was synthesized by the solid-state reaction as described in Ref. [17,19]. Next, we charged the powder of $VOSe_2O_5$ (0.5 g) with the transport agent $NH_4Cl$ (around 30 mg) at an end of the sealed quartz tube (length: 120 mm, diameter: 18 mm), and placed this tube in a three-zone furnace with the temperature gradient from 400 °C to 350 °C. The duration for the main transport process is typically two months. We introduced a pre-reaction process with the temperature 380 °C on the source side for around four days in order to induce the partial decomposition reaction: $VOSe_2O_5 \rightarrow VOSeO_3 + SeO_2$, which is important to eliminate the production of $VO_2$ on the source side, as $VO_2$ is detrimental for the formation of $VOSe_2O_5$ as the final product.

Figure S1(a) shows the photograph of the co-aligned single crystals for the SANS measurements. Each sample has clear crystallographic facets typically for (001), (100), and (110) planes. The co-alignment was performed by putting the samples on the flat aluminum plate, and attaching a facet on one crystal with that of the other, for example, one's (100) plane with the other's (-100) plane (see Fig. S1(b)). For the in-plane orientation, we performed the single crystal x-ray diffraction for each sample in order to distinguish (100) planes from (110) planes.

The demagnetization field for the co-aligned sample shifts the magnetic phases towards higher field than those of the single piece of a crystal. We performed the measurement of magnetic properties for the assembled sample as shown in Figs. S1(c)-S1(f)). Figure S1(c) shows the temperature dependence of $\chi'$ in zero field with the ac magnetic field ($H_{ac}$) along the $c$ axis. The onset of the cycloidal spin state (IC-1) is clearly discerned by a peak at 7.50 K, and the transition to the B phase is also observed as indicated by a black triangle. These clear phase transitions indicate the homogeneous high quality of each single crystal for the SANS measurement. Figure S1(d) shows the behavior of $\chi'$ around $T_C$ under $H//c$. For $H$ = 20 Oe ~ 30 Oe, a dip structure is observed, indicating the triangular SkL state. The SkL state is also confirmed in the $H$ dependence of $\chi'$ as shown in Fig. S1(e). Clear double-peak structure is observed at 7.45 K. This feature is slightly broadened as decreasing temperature due perhaps to the inhomogeneous demagnetization field. This may be the reason for the multidomain state in terms of the triangular SkL, IC-1, and IC-2 states as seen in Fig. 3(a). As indicated by a black triangle in Fig. S1(e), shoulder structure is discernible. We assigned it to the phase boundaries for the triangular SkL state against the IC-1 phase and FM phase with spins parallel to the $c$ axis. Figure S1(f) is the $H$-$T$ phase diagram with $H//c$ determined by the magnetization measurements with the assembled-crystals sample. The magnetic fields for the phase boundaries are shifted towards high field due to the demagnetization field, but the clear pocket-like region for the triangular SkL state is identified.

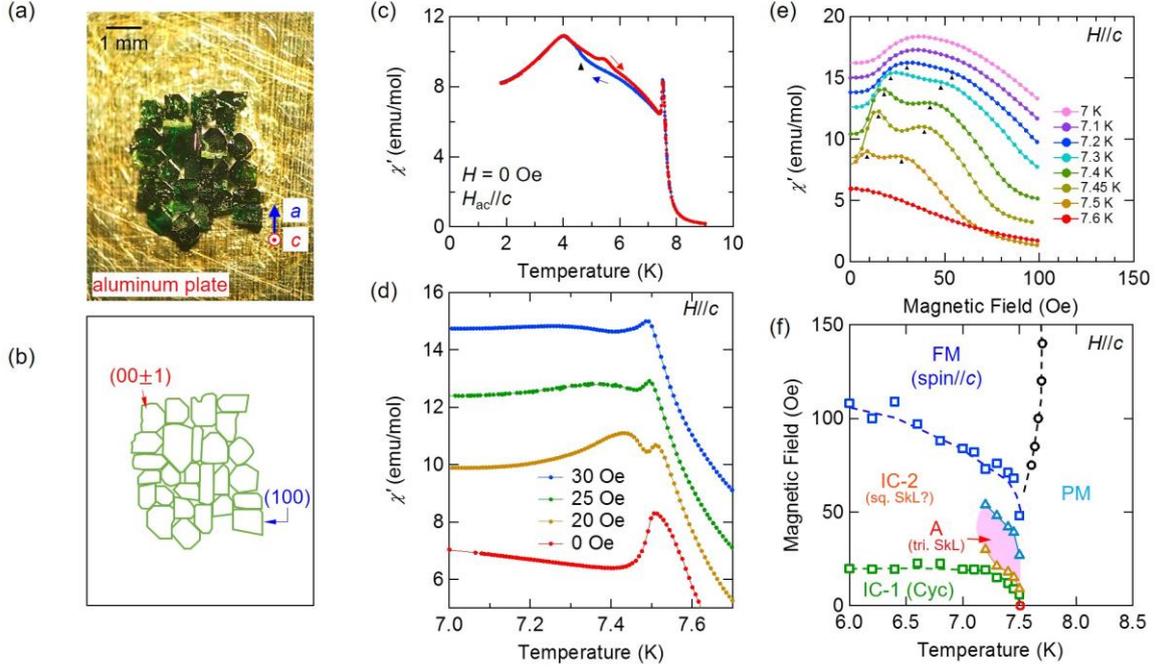

FIG. S1. (a) Photograph of the sample for the SANS measurement. Twenty-seven pieces of single crystals are assembled to be co-aligned with the *a* and *c* axes. (b) Outline of each single crystal. (c-e) The real part of the ac magnetic susceptibility $\chi'$ data for the assembled single crystals. (c) Temperature dependence of $\chi'$ in zero field. Black triangle indicates the anomaly for the transition from IC-1 phase to the B phase in the cooling process. (d,e) Temperature and *H* dependence of $\chi'$ with *H*//*c*, respectively. Data are shifted vertically for clarity. In (e), black triangles indicate the peak or shoulder structures of $\chi'$, assigned to the phase boundary between triangular SkL state and IC-1/FM state. (f) *H-T* phase diagram under *H*//*c* for the assembled single crystals determined by the magnetization measurements.

## 2. SANS investigation with a magnetic field along the *a* axis

SANS measurements were performed using the instrument KWS-1 of Jülich Centre for Neutron Science (JCNS) at Heinz Maier-Leibnitz Zentrum (MLZ), Garching, Germany [21,22]. The neutron beam was collimated over a length of 20 m before reaching the sample. The scattered neutrons were counted by a two-dimensional (2D) position-sensitive detector located at 20 m downstream of the sample. The neutron wavelength was set to 10 Å. To measure the SANS pattern for the beam along the *c* axis, twenty-seven pieces of crystals, of the total volume of 4.6 mm$^3$, were aligned on an aluminum plate by their (001) crystal surface and in-plane orientation co-aligned by X-ray diffraction. To maximize the signal to background ratio a boron-carbide mask was mounted directly on the sample holding Al-plate around the sample. The sample on the Al-plate was mounted into a $^3$He-circulation refrigerator with its (001) direction parallel to the incident beam. The precision of the temperature is better than ±0.01 K during each integration time. For the case of the field-scan measurement under *H*//*a* (Figs. 1(e)-1(f) and Figs. S2(a)-S2(e)), the temperature gradually increases from 6.18 K (Fig. 1(e)) to 6.27 K (Fig. 1(f)), where the temperature fluctuation for each measurement is 0.003 K on average and 0.009 K in maximum. The incident beam was narrowed by the built-in diaphragm to the co-aligned-crystals sample area of 6 × 4 mm$^2$ on the sample. A magnetic field along or perpendicular to the incident beam was generated by an electro-magnet. To avoid the effect of the residual field, the electromagnet was demagnetized before performing the zero and subsequent applied field experiments. Background was measured in the field-polarized state of the sample and subtracted from the SANS patterns.

Figures S2(a)-S2(e) show the field-evolution of the SANS pattern for the co-aligned single crystals under *H*//*a* (see Fig. S2(e) for the direction of *H*). The neutron beam is parallel to the *c* axis. Four-fold pattern in zero field (Fig. S2(a)) indicates the multidomain state of the single-*q* cycloidal spin order. On the other hand for *H* = 120 Oe (Fig. S2(e)), no magnetic scattering is discerned due to entering the field-forced ferrimagnetic state as shown in Fig. 1(d). Prior to the transition to ferrimagnetic state,

we observed that the vertical spots for $\vec{q} \parallel \vec{a}$ rapidly disappear as the field increases while the intensity of the horizontal spots for $\vec{q} \parallel \vec{b}$ remains. The two-fold pattern for $H$ = 80 Oe in Fig. S2(d) indicates that the single domain state of the single-$q$ cycloidal spin state is selected.

This field-induced domain rearrangement is consistent with the hysteresis of magnetic property under $H//a$. Figures S2(f)-S2(g) show the $H$ dependence of magnetic susceptibility $\chi'$ and the integrated intensities, $I_v$ and $I_h$, for the vertical and horizontal sectors, respectively. The integration region for each component is shown as the inset in Fig. S2(g). The hysteresis for $\chi'$ closes at around 20 Oe, which is consistent with the disappearance of the $I_v$ in the $H$-increasing process. The $I_v$ does not recovered when $H$ goes back to the zero field from 120 Oe, while the $I_h$ slightly gains as compared to the initial value. This corresponds to the two-fold pattern shown in Fig. 1(f), in which a single-$q$ domain in $q \perp H$ condition is selected by the field-trained process.

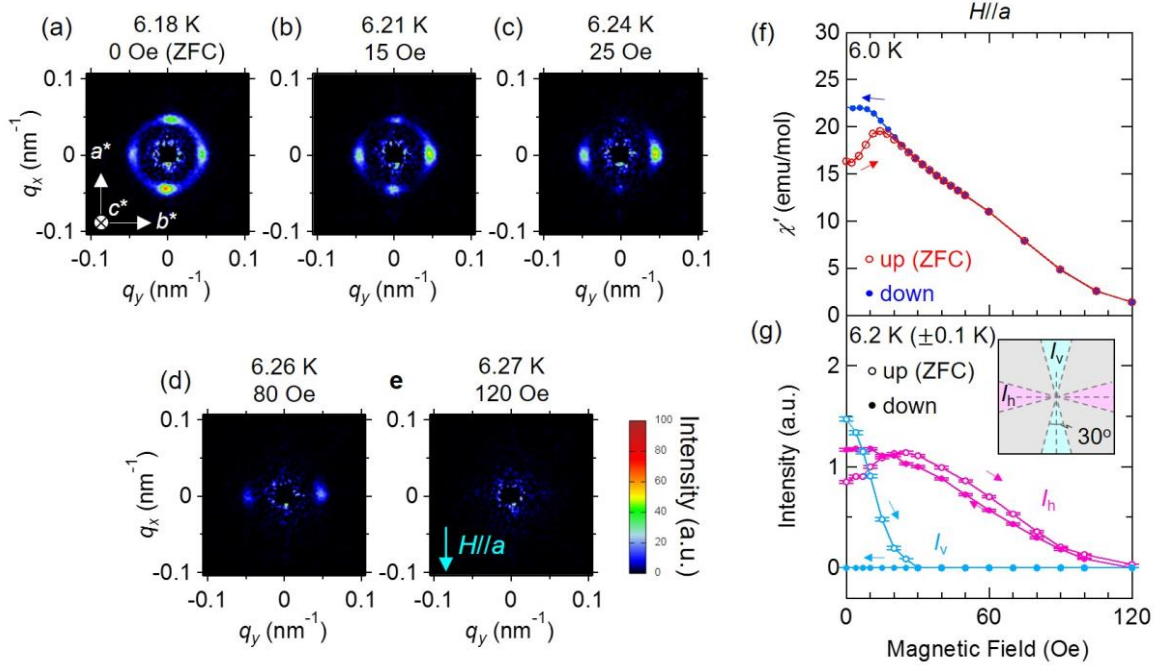

FIG. S2. (a-e) Magnetic-field ($H$) induced evolution of the SANS pattern for the co-aligned single crystals under $H//a$. The direction of $H$ is indicated in (e). The measurement was performed in the $H$-increasing process at 6.2 K ($\pm 0.1$ K). During the measurement, the temperature gradually increases from 6.18 K (Fig. 1(e)) to 6.27 K (Fig. 1(f)), where the temperature fluctuation for each measurement is 0.003 K on average and 0.009 K in maximum. Prior to the measurement, the sample was cooled down in zero field from the paramagnetic region. (f,g) $H$ dependence of (f) the real part of the ac magnetic susceptibility $\chi'$, and (g) the integrated SANS intensities, $I_v$ (blue circle) and $I_h$ (pink circle), involving the Bragg reflections for the vertical and the horizontal region, respectively. The integration regions for $I_v$ and $I_h$ are the blue and pink region, respectively, as shown in the inset. The measurement was performed after the zero-field cooling process followed by the $H$-increasing scan (open circle), and then $H$-decreasing scan (closed circle).